\documentclass[a4paper,11pt]{article}

\usepackage[utf8]{inputenc}
\usepackage[T1,T2A]{fontenc}
\usepackage{amsmath,amsfonts,amssymb}

\usepackage{cite}

\usepackage{xcolor}

\usepackage{graphicx}
\usepackage{wrapfig}
\usepackage{hyperref}

\usepackage{geometry}
\numberwithin{equation}{section}

\newcommand{\lb}[0]{\left(}
\newcommand{\rb}[0]{\right)}
\newcommand{\lsb}{\left[}
\newcommand{\rsb}{\right]}
\newcommand{\pz}{\partial_z}
\newcommand{\veps}{\varepsilon}

\begin{document}

\renewcommand*{\thefootnote}{\fnsymbol{footnote}}

\begin{center}
{\Large\bf Cornell potential in generalized Soft Wall holographic model}
\end{center}
\bigskip
\begin{center}
{Sergey Afonin\footnote{E-mail: \texttt{s.afonin@spbu.ru}.}
and Timofey Solomko}
\end{center}

\renewcommand*{\thefootnote}{\arabic{footnote}}
\setcounter{footnote}{0}

\begin{center}
{\small\it Saint Petersburg State University, 7/9 Universitetskaya nab.,
St.Petersburg, 199034, Russia}
\end{center}

\bigskip

\begin{abstract}
We derive and analyze the confinement potential of the Cornell type within the framework of the generalized Soft Wall holographic model
that includes a parameter controlling the intercept of the linear Regge spectrum.
In the phenomenology of Regge trajectories, this parameter is very important for the quantitative description of experimental data.
Our analysis shows that the ``linear plus Coulomb'' confinement potential
obtained in the scalar channel is {\it quantitatively} consistent with the phenomenology and lattice simulations
while the agreement in the vector channel is qualitative only.
This result indicates the key role of the vacuum scalar sector in the formation of the confinement potential.
As a by-product the overall consistency of our holographic description
of confinement potential seems to confirm the glueball nature of the scalar meson $f_0(1500)$.
\end{abstract}



\bigskip
\bigskip

\section{Introduction}

The quark confinement remains an unresolved key problem of strong interactions despite
a tremendous number of works devoted to this subject.  One of the basic observables relevant
to confinement is the heavy-quark potential. In the real life, this potential is saturated at a constant level
at large enough (of the order of 1~fm) distances because of the light quark-antiquark pairs
popping up out of the vacuum and thereby completely screening the static sources.
But the approximation of heavy static quarks allows to simplify the problem and to address it
both analytically within various phenomenological models and in lattice simulations
directly from QCD. For this reason, the limit of heavy static quarks is interesting and informative.

The detailed lattice simulations of the form of the heavy-quark potential (see, e.g.,
the review~\cite{bali}) revealed a remarkable agreement
with the Cornell potential~\cite{Eichten:1978tg},
\begin{equation}
\label{cornell}
V(r)=-\frac{\kappa}{r}+\sigma r + \text{const}.
\end{equation}
This result imposes a serious restriction on viable phenomenological approaches to strong interactions:
In the non-relativistic limit, they should be able to reproduce the behavior~\eqref{cornell}.

One of such promising approaches that passes the given test is the so-called Soft-Wall (SW)
holographic model~\cite{son2,andreev}. This popular bottom-up AdS/QCD approach was originally inspired by the
AdS/CFT correspondence in string theory~\cite{mald,witten,gub} and turned out to be unexpectedly
successful in the description of hadron Regge spectroscopy, hadron form-factors, QCD thermodynamics,
and other phenomenology related to the non-perturbative strong interactions (many contemporary
references are collected in Ref.~\cite{Afonin:2021cwo}).
The heavy-quark potential was first calculated within the SW holographic model by Andreev and Zakharov in
Ref.~\cite{Andreev:2006ct} and their result agreed with~\eqref{cornell}. The analysis in Ref.~\cite{Andreev:2006ct},
however, was performed only for one particular case of the SW model: The vector model with a fixed (simplest)
intercept of string like mass spectrum, $m^2_n=an$, where $n=1,2,\dots$.
In view of high citation of this analysis in the literature, it is interesting
to extend it to arbitrary intercept of linear radial Regge spectrum, $m^2_n=a(n+b)$, and to the case of scalar SW model.
To the best of our knowledge, such extensions were not considered in the literature. The purpose of our present work
is to fill this gap. Since the intercept $b$ parametrizes important effects of low-energy strong interactions
and is indispensable for making quantitative phenomenological predictions, the holographic derivation and analysis of
heavy-quark potential in the presence of nontrivial intercept should provide an interesting test for overall
phenomenological consistency of the approach.

The linear confinement potential was also recovered in a closely related AdS/QCD approach called the light-front holographic QCD~\cite{br3}
and a good numerical agreement was observed, see Refs.~\cite{Brodsky:2010ur,Trawinski:2014msa} for the corresponding discussions. A similar conclusion will be made in
our work: We will demonstrate that the generalized SW holographic model leads to a good quantitative description of the shape of
the Cornell potential~\eqref{cornell}.

The paper is organized as follows. In Section~2, the general design of generalized SW holographic model is presented.
The derivation of potential energy between static sources from Holographic Wilson loop within the generalized vector SW model is given in Section~3.
The large and small distance asymptotics of this energy are calculated in Sections~4 and~5, respectively.
Some technical details are transferred to the Appendices. In Section~6,
we extend the results to the generalized scalar SW holographic model. The phenomenological predictions are discussed in Section~7.
We conclude in Section~8.

\section{Generalized Soft Wall holographic model}

The standard SW holographic model is defined by the 5D action~\cite{son2}
\begin{equation}
\label{sw}
  S=\int d^4\!x\,dz\sqrt{g}\,e^{-cz^2}\mathcal{L},
\end{equation}
where $g=|\text{det}g_{MN}|$, $\mathcal{L}$ is a Lagrangian density of some
free fields in AdS\(_5\) space which, by assumption, are
dual on the AdS\(_5\) boundary to some QCD operators.
The metric is given by the Poincar\'{e} patch of the AdS$_5$ space,
\begin{equation}
\label{2}
g_{MN}dx^Mdx^N=\frac{R^2}{z^2}(\eta_{\mu\nu}dx^{\mu}dx^{\nu}-dz^2),\qquad z>0.
\end{equation}
Here $\eta_{\mu\nu}=\text{diag}(1,-1,-1,-1)$, $R$ denotes the radius of AdS$_5$ space,
and $z$ is the holographic coordinate which has the standard physical interpretation of the inverse energy scale.
The static dilaton background $e^{-cz^2}$ (with $c>0$ in the original model of Ref.~\cite{son2})
gives rise to Regge behavior of mass spectrum. The standard SW model is defined in the probe
approximation, i.e., the metric is not backreacted by matter fields and dilaton
--- such backreaction
is assumed to be suppressed in the large-$N_c$ limit (it should be recalled that, strictly speaking,
the holographic approach is formulated in this limit only).
There are many bottom-up holographic models based on Einstein-dilaton gravity that provide backgrounds
consistent with Einstein equations. This line of research was pioneered in the work~\cite{Gursoy:2007er},
the most recent review is presented in~\cite{Chen:2022goa}.
The great advantage of the probe approximation is its simplicity which allows analytical control at each stage.
In relation to our task, extracting the confinement potential beyond this approximation would be a hard technical problem.
On the other hand, we will argue that our approach leads (effectively) to a certain phenomenological model for spin-dependent
backreaction to the AdS metric by an injected particle.

The Lagrangian density of the simplest SW model describing vector mesons is~\cite{son2}
\begin{equation}
\label{1}
\mathcal{L}=-\frac{1}{4}F_{MN}F^{MN}+\frac12m_5^2V_MV^M,
\end{equation}
where $F_{MN}=\partial_M V_N-\partial_N V_M$, $M,N=0,1,2,3,4$.
According to the standard prescriptions of AdS/CFT correspondence~\cite{witten,gub},
the 5D mass $m_5$ is determined by the behavior of $p$-form fields near the UV boundary $z=0$,
\begin{equation}
\label{3}
m_5^2R^2=(\Delta-p)(\Delta+p-4),
\end{equation}
where $\Delta$ denotes the scaling dimension of 4D operator dual to the corresponding 5D field on the UV boundary.
In the vector case $p=1$, thus $m_5^2R^2=(\Delta-1)(\Delta-3)$.
The canonical dimension of vector current operator in QCD is $\Delta=3$ that leads to $m_5=0$.
This corresponds to massless 5D vector fields which are usually considered in the SW models.

The 4D mass spectrum of the model can be found, as usual, from the equation of
motion accepting the 4D plane-wave ansatz $V_\mu(x,z)=e^{ipx}v(z)\epsilon_\mu$
with the on-shell, $p^2=m^2$, and transversality, $p^\mu\epsilon_\mu=0$, conditions.
Here, \(v(z)\) is a profile function for physical 4D modes. In addition, the condition
$V_z=0$ is implied for the physical components of 5D fields~\cite{br3}. For massless
vector fields, this is equivalent to the standard choice of axial gauge due to emerging
gauge invariance~\cite{son2}. The equation of motion ensuing from the action~\eqref{1}
for $m_5=0$ is
\begin{equation}
\label{4}
\partial_z\left(\frac{e^{-cz^2}}{z}\partial_z v_n\right)=-\frac{m_n^2}{z}e^{-cz^2}v_n,
\end{equation}
The particle-like excitations correspond to normalizable solutions of Sturm-Liouville
equation~\eqref{4}, which are enumerated by the index \(n\). It is convenient to make
the substitution
\begin{equation}
\label{5}
v_n=z^{1/2}e^{cz^2/2}\psi_n,
\end{equation}
that transforms the Eq.~\eqref{4} into a form of one-dimensional Schr\"{o}dinger equation
\begin{equation}
\label{6}
-\partial_z^2\psi_n+V(z)\psi_n=m_n^2\psi_n,
\end{equation}
with the potential
\begin{equation}
\label{7}
V(z)=c^2z^2+\frac{3}{4z^2}.
\end{equation}
The mass spectrum of the model is given by the eigenvalues of Eq.~\eqref{6},
\begin{equation}
\label{8}
m_n^2=4|c|n,\qquad n=1,2,\dots.
\end{equation}

There exists another formulation of SW model proposed independently and almost simultaneously in Ref.~\cite{andreev},
in which the dilaton background is absent but the AdS$_5$ metric is modified,
\begin{equation}
\label{az_metric}
  g_{MN}=\text{diag}\left\lbrace\frac{R^2}{z^2}h,\dots,\frac{R^2}{z^2}h\right\rbrace,\quad
  h=e^{-2cz^2}.
\end{equation}
The resulting equation of motion and spectrum are the same.
Other formulations of SW holographic model are also possible. They are analyzed in detail
in the recent work~\cite{Afonin:2021cwo}.

The generalized SW holographic model describes the linear spectrum with arbitrary intercept regulated by a parameter $b$.
In the vector case, the generalization of spectrum~\eqref{8} is
\begin{equation}
\label{spSW}
m_n^2=4|c|(n+b),\qquad n=1,2,\dots.
\end{equation}
As was shown in~\cite{Afonin:2012jn} and further developed in~\cite{Afonin:2021cwo},
the spectrum~\eqref{spSW} arises in the following generalization of the action of vector SW model,
\begin{equation}
\label{gen_sw}
  S=\int d^4xdz\sqrt{g}e^{-cz^2}U^2(b,0,|cz^2|)\mathcal{L},
\end{equation}
where $U$ is the Tricomi hypergeometric function that modifies the dilaton background.
It should be remarked that any other modification of quadratic dilaton background, either manually
or by considering a backreacted geometry, will lead to a distortion of the exact Regge behavior,
$m_n^2\sim n$, for any $n$~\cite{Afonin:2021cwo} (and can even lead to a finite number of excitations~\cite{Afonin:2009xi}).

The generalized SW model can be also reformulated in the form of the modified metric~\eqref{az_metric}.
Generically, if a 5D holographic model describing spin-$J$ mesons contains a $z$-dependent background $B(z)$ in its action,
\begin{equation}
\label{ac1}
  S=\int d^4x dz\sqrt{g}\,B(z)\,\mathcal{L},
\end{equation}
this action can be rewritten in the form
\begin{equation}
\label{ac2}
  S=\int d^4x dz\sqrt{\tilde{g}}\,\tilde{\mathcal{L}},
\end{equation}
with the modified metric~\cite{Afonin:2021cwo}
\begin{equation}
\label{tr}
  \tilde{g}_{MN}=B^{(3/2-J)^{-1}}g_{MN}.
\end{equation}
Substituting the background $B(z)$ in the action~\eqref{gen_sw} for spin $J=1$, we obtain the following generalization
for the function $h$ in the modified metric~\eqref{az_metric},
\begin{equation}
\label{gen_h}
  h=e^{-2cz^2}U^4(b,0,|cz^2|).
\end{equation}
This modification will be the starting point for our further analysis of Cornell potential within the generalized SW holographic model.

A question may appear how should we interpret the result~\eqref{tr} which would imply that there could be different 5D metrics for fields of
different spin? On the one hand, this is a formal consequence of rewriting of the action~\eqref{ac1} in the form of the action~\eqref{ac2}.
The first action is for fields which describe particles of
arbitrary integer spin moving in the AdS space with universal $z$-dependent dilaton background. The second action describes particles
moving in a modified AdS space without background but the modification of AdS metric becomes spin-dependent. The modification of metric is such that
the harmonic oscillatory part of the potential~\eqref{7} in the equation of motion remains universal for any spin and this provides the universal
slope for Regge trajectories (which is an important manifestation of confinement). On the other hand, the second formulation of
generalized SW model becomes primary for analysis of confinement properties, thus there should be some physics behind such a mathematical
reformulation. We may give the following qualitative physical meaning to the second description.
In a space with confining geometry, particle moves around the minimum of its effective gravitational potential energy~\cite{br3}.
But part of this effective gravitational potential can arise dynamically due to the backreaction of the particle mass
(that depends on the particle spin) to the geometry of the environment. The relation~\eqref{tr} can be then interpreted as a phenomenological
model for this backreaction.

Concluding the reminder we would make the following remark. The linear Regge like spectrum is usually interpreted as a manifestation
of confinement. Within the Light-Front holographic QCD~\cite{br3}, the holographic coordinate $z$ has the physical meaning of the measure of
the distance between a quark and an antiquark. Then $z$ is proportional to $r$ in the Cornell potential~\eqref{cornell}. In this case,
the holographic potential~\eqref{7} can be interpreted, up to an additive constant, as the square of the real potential energy~\eqref{cornell} between static quarks.
The form of Cornell potential~\eqref{cornell} is then hinted in~\eqref{7}. On the heuristic level, if the energy $E$ in the
Schr\"{o}dinger equation is replaced by the energy squared $E^2$ (as is suggested by the Lorentz invariance in the static limit),
then the oscillator potential leads to the behavior $E^2\sim r^2$ at large distance $r$, i.e., to the linearly rising potential
energy $E\sim r$. In a sense, it is this situation that is encoded in the SW holographic model. For the case of Light-Front holographic QCD,
this point is discussed in detail in Ref.~\cite{Trawinski:2014msa}.

\section{Holographic Wilson loop}

\begin{wrapfigure}{r}{0.4\textwidth}
  \vspace{-15mm}
  \begin{center}
    \includegraphics[width=0.38\textwidth]{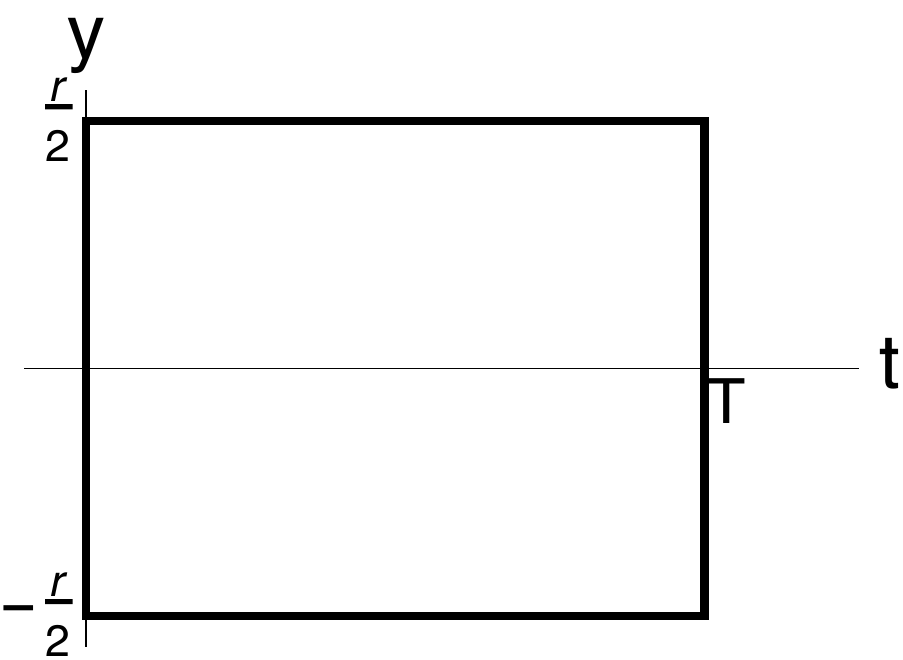}
  \end{center}
  \vspace*{-7mm}
  \caption{A Wilson loop.}
  \vspace*{-5mm}
  \label{wilson_loop}
\end{wrapfigure}

The analytic and lattice calculations of the potential between static sources are usually based on
the analysis of a Wilson loop. The holographic variant of this analysis was developed by Maldacena
in Ref.~\cite{Maldacena:1998im}. Within the holographic framework, one considers
a Wilson loop \(\mathcal{L}\) placed in the 4D boundary of the 5D space with the
time coordinate ranging from \(0\) to \(T\) and the remaining 3D spatial coordinates \(y\)
from \(-r/2\) to \(r/2\), see Figure~\ref{wilson_loop}. The expectation value of the loop
in the limit of \(T\to\infty\) is as usual
\begin{equation}
  \left\langle W(\mathcal{C})\right\rangle\sim e^{-TE(r)},
\end{equation}
where \(E(r)\) is the energy of the quark-antiquark pair. Alternatively, this expectation
value can be obtained via
\begin{equation}
  \left\langle W(\mathcal{C})\right\rangle\sim e^{-S},
\end{equation}
where \(S\) represents the area of a string world-sheet which produces the loop \(\mathcal{L}\).
Combining these two equations one can compute the energy (the static potential) of configuration
as
\begin{equation}\label{en_def}
  E=\frac{S}{T}.
\end{equation}
The natural choice for the world-sheet area is the Nambu-Goto action
\begin{equation}\label{ng}
  S=\frac{1}{2\pi\alpha'}\int d^2\xi\sqrt{\det g_{MN}\partial_\alpha X^M\partial_\beta X^N},
\end{equation}
where \(\alpha'\) is the inverse string tension, \(X^M\) are the string coordinates
functions which provide a mapping from the \(\lb\xi_1,\xi_2\rb\) parameter space of the
world-sheet into the spacetime, and \(g_{MN}\) is the modified Euclidean AdS metric
\begin{equation}\label{metric}
  g_{MN}=\text{diag}\left\lbrace\frac{R^2}{z^2}h,\dots,\frac{R^2}{z^2}h\right\rbrace.
\end{equation}
The background function \(h(z)\) specifies a holographic model,
the general requirement is that the metric must be asymptotically AdS at \(z\to0\). After choosing
\(\xi_1=t\) and \(\xi_2=y\) as the parametrization of the world-sheet and integration
over \(t\) from \(0\) to \(T\), the action can be rewritten as
\begin{equation}\label{ng_action}
  S=\frac{TR^2}{2\pi\alpha'}\int\displaylimits_{-r/2}^{r/2}dy\,
  \frac{h}{z^2}\sqrt{1+z'^2},
\end{equation}
where \(z'=dz/dy\). From the first integral of this action, which corresponds to the
action's translational invariance, one can then obtain an integral expression for the
distance \(r\),  the end result is
\begin{equation}\label{r_int_def}
  r=2\sqrt{\frac{\lambda}{c}}\int\displaylimits_{0}^{1}dv\,\frac{h_0}{h}
  \frac{v^2}{\sqrt{1-v^4\frac{h_0^2}{h^2}}},
\end{equation}
where we introduced a new notation,
\begin{equation}\label{not}
  z_0\equiv\left.z\right|_{y=0},\quad
  h_0\equiv\left.h\right|_{z=z_0},\quad
  v\equiv\frac{z}{z_0},\quad
  \lambda\equiv cz_0^2.
\end{equation}

The expression for the energy is obtained from the equation~\eqref{en_def} and the
action~\eqref{ng_action} by replacing the integration over \(y\) with the integration over
\(v\) by using~\eqref{r_int_def} (note that \(r\) is equal to the integral over \(dy\)
from \(-r/2\) to \(r/2\)). The end result is
\begin{equation}
\label{E2}
  E=\frac{R^2}{\pi\alpha'}\sqrt{\frac{c}{\lambda}}\int\displaylimits_{0}^{1}\frac{dv}{v^2}\,
  \frac{h}{\sqrt{1-v^4\frac{h_0^2}{h^2}}}.
\end{equation}
The details of these derivations can be found, e.g., in Ref.~\cite{Afonin:2021zdu}.

This integral in~\eqref{E2} is evidently divergent at \(v=0\) due to the \(v^2\) in the denominator of
the integrand. To solve this problem we introduce a regularization by imposing a cutoff
\(\veps\to 0\)
\begin{equation}
  E=\frac{R^2}{\pi\alpha'}\sqrt{\frac{c}{\lambda}}\int\displaylimits_{0}^{1}
  \frac{dv}{v^2}\lb\frac{h(\lambda,v)}{\sqrt{1-v^4\frac{h_0^2}{h^2}}}-D\rb+
  \frac{R^2}{\pi\alpha'}\sqrt{\frac{c}{\lambda}}D
  \int\displaylimits_{\veps/z_0}^{1}\frac{dv}{v^2},
\end{equation}
where the regularization constant is defined as
\begin{equation}
  D\equiv \left.h\right|_{v=0}.
\end{equation}
The second integral can be easily computed (here, we temporarily switch notation back to
\(1/z_0=\sqrt{c/\lambda}\))
\begin{equation}\label{co_int_res}
  \frac{R^2}{\pi\alpha'}\frac{D}{z_0}\int\displaylimits_{\veps/z_0}^1\frac{dv}{v^2}=
  -\frac{R^2}{\pi\alpha'}\frac{D}{z_0}\left.\frac{1}{v}\right|_{\veps/z_0}^1=
  \frac{R^2}{\pi\alpha'}\frac{D}{z_0}\lb\frac{z_0}{\veps}-1\rb=
  \frac{R^2D}{\pi\alpha'\veps}-\frac{R^2}{\pi\alpha'}\frac{D}{z_0}.
\end{equation}
Then we introduce the \textit{regularized energy}
\begin{equation}
  E_R=\frac{R^2D}{\pi\alpha'\veps}+E,
\end{equation}
with the corresponding redefinition of the energy itself as
\begin{equation}\label{en_int_def}
  E=\frac{R^2}{\pi\alpha'}\sqrt{\frac{c}{\lambda}}\lsb\int\displaylimits_{0}^{1}
  \frac{dv}{v^2}\lb\frac{h}{\sqrt{1-v^4\frac{h_0^2}{h^2}}}-D\rb-D\rsb.
\end{equation}

Consider now the case of generalized vector SW holographic model.
According to the discussions in Section~2, the background function
is given by the expression~\eqref{gen_h}.
The background used by Andreev and Zakharov in Ref.~\cite{Andreev:2006ct} had the opposite sign in the
exponent to enforce a confining geometry.
We will use the result~\eqref{gen_h} to generalize the Andreev-Zakharov deformed AdS metric, i.e.,
we will investigate the confinement properties of SW model with the following generalized
background function ($c>0$),
\begin{equation}
\label{bg_f}
  h=e^{2cz^2}U^4(b,0,cz^2) = e^{2\lambda v^2}U^4(b,0,\lambda v^2).
\end{equation}
Substituting~\eqref{bg_f} into the integrals for the distance~\eqref{r_int_def} and the energy~\eqref{en_int_def}
we get
\begin{equation}\label{r_int_def_2}
  r=2\sqrt{\frac{\lambda}{c}}\int\limits_0^1dv\,
  \frac{U^4(b,0,\lambda)}{U^4(b,0,\lambda v^2)}
  \frac{v^2e^{2\lambda(1-v^2)}}
  {\sqrt{1-v^4e^{4\lambda(1-v^2)}\frac{U^8(b,0,\lambda)}{U^8(b,0,\lambda v^2)}}},
\end{equation}
\begin{equation}\label{en_int_def_2}
  E=\frac{R^2}{\pi\alpha'}\sqrt{\frac{c}{\lambda}}\lsb\int\limits_0^1\frac{dv}{v^2}\lb
  \frac{e^{2\lambda v^2}U^4(b,0,\lambda v^2)}{\sqrt{1-v^4e^{4\lambda(1-v^2)}\frac{U^8(b,0,\lambda)}{U^8(b,0,\lambda v^2)}}}-D\rb-D\rsb,
\end{equation}
where the regularization constant is \(D=U^4(b,0,0)\).

The last two expressions determine the energy as a function of the distance in the parametric form. Our
next goal is to obtain small and large distance behavior of the energy. In
order to derive the corresponding \(E(r)\) asymptotics we have to map them into corresponding
\(\lambda\) asymptotics of \(r\) and \(E\). We will first focus on the large distance
asymptotics.

\section{Large distance potential}


Since the potential \(E(r)\) is defined in a parametric form, one has to establish a
correspondence between large distances and the range of values of the parameter \(\lambda\)
via the analysis of the integrals~\eqref{r_int_def_2}
and~\eqref{en_int_def_2}. Note that these integrals must be real-valued, and as such the
expression under the square root must be greater than zero within the integration range,
\(v\in\lsb 0,1\rsb\),
\begin{equation}
\label{sqrt_exp}
  1-v^4e^{4\lambda(1-v^2)}\frac{U^8(b,0,\lambda)}{U^8(b,0,\lambda v^2)}>0.
\end{equation}
This can be satisfied if the minimum value of the expression~\eqref{sqrt_exp} is strictly positive (on the
same interval), thus, we have to analyze the roots of the derivative of expression~\eqref{sqrt_exp},
\begin{equation}
\label{eq12}
  1-2\lambda v^2-4\lambda v^2\dfrac{U'(b,0,\lambda v^2)}{U(b,0,\lambda v^2)}=0, \qquad
  U(b,0,\lambda v^2)\ne 0,
\end{equation}
where \(U'(b,0,x)=\partial_x U(b,0,x)\). Denoting
\begin{equation}
\label{eq12b}
  x\equiv\lambda v^2,
\end{equation}
the numerical calculations show that the integrals are real-valued if
\begin{equation}
  \lambda<x.
\end{equation}
The same conditions can be also derived from the so-called Sonnenschein
conditions~\cite{Sonnenschein:2000qm} which state that the \(g_{00}\) element of the metric
must satisfy
\begin{equation}\label{sonn_cond}
  \left.\pz g_{00}\right|_{z=z_0}=0,\quad
  \left.g_{00}\right|_{z=z_0}\ne0,
\end{equation}
in order for the background to be dual to a confining theory in the sense of the area law
behavior of a Wilson loop.

The integral~\eqref{r_int_def_2} is the growing function of \(\lambda\), hence, the
large distances correspond to large values of the parameter \(\lambda\). We get from above that
technically the limit \(\lambda\to x\) should be examined in the derivation of the large distance asymptotics.
Our procedure includes the following steps. First, we note that the main contribution to the
integrals comes from the upper integration bound, \(v=1\), since the integrals diverge
in the upper limit. Hence, we should expand the integrands around that point or, rather, the expression
under the square root, since other factors under the integral either do not diverge or
are constant functions of \(v\). This leads to the expressions,
\begin{equation}\label{rbeh}
  r\underset{v\to1}=2\sqrt{\frac{\lambda}{c}}\int\limits_0^1\frac{dv}{\sqrt{A(b,\lambda)(1-v)+B(b,\lambda)(v-1)^2}},
\end{equation}
\begin{equation}\label{Ebeh}
  E\underset{v\to1}=\frac{R^2}{\pi\alpha'}\sqrt{\frac{c}{\lambda}}e^{2\lambda}U^4(b,0,\lambda)
  \int\limits_0^1\frac{dv}{\sqrt{A(b,\lambda)(1-v)+B(b,\lambda)(v-1)^2}},
\end{equation}
where, most notably, the remaining integrals are exactly the same due to identical expressions
under the square root. Here the functions \(A(b,\lambda)\) and \(B(b,\lambda)\) are
\begin{equation}
  A(b,\lambda) = -s'_v(b,\lambda;1)=
  4\lsb 1-2\lambda-4\lambda\frac{U'(b,0,\lambda)}{U(b,0,\lambda)}\rsb,
\end{equation}
\begin{multline}
  B(b,\lambda) = \frac{s''_{vv}(b,\lambda;1)}{2}=\\-2\lsb 16\lambda^2-18\lambda+3+
  72\lambda^2\frac{U'(b,0,\lambda)^2}{U(b,0,\lambda)^2}+
  4\lambda\frac{(16\lambda-9) U'(b,0,\lambda)-2 \lambda U''(b,0,\lambda)}{U(b,0,\lambda)}\rsb,
\end{multline}
where
\begin{equation}
  s(b,\lambda;v)\equiv 1-v^4e^{4\lambda(1-v^2)}\frac{U^8(b,0,\lambda)}{U^8(b,0,\lambda v^2)}.
\end{equation}

In the second step, we express the integral in~\eqref{rbeh} via $r$ and substitute it into~\eqref{Ebeh}.
Finally, we take the limit of \(\lambda\to x\) in the found expression and obtain the asymptotics under consideration,
\begin{equation}
\label{pot_large_r}
  E\underset{r\to\infty}{=}\frac{R^2}{\alpha'}\sigma_\infty r,
\end{equation}
\begin{equation}
\label{large_r_not}
  \sigma_\infty=\frac{e^{2x}U^4(b,0,x)}{2\pi x}c.
\end{equation}
As is clear from our discussions above, here $x=x(b)$ represents a non-trivial function of $b$
defined by the Eq.~\eqref{eq12}.
In the case of the standard SW model, $b=0$, one has $x(0)=1/2$ and the asymptotics~\eqref{large_r_not}
reduces to the corresponding result of Andreev and Zakharov's analysis in Ref.~\cite{Andreev:2006ct},
\begin{equation}
\label{large_r_not_0}
  \sigma_\infty(0)=\frac{e}{\pi}c\approx 0.87 c,
\end{equation}
the relevant details are discussed in the Appendix~\ref{adj_app}.
It is known, however, that the simple case of $b=0$ in~\eqref{spSW} does not reproduce correctly neither the
radial spectrum of light vector mesons nor the pion form-factor. The phenomenology clearly suggests
a value of intercept near (see the detailed discussions on this point in Ref.~\cite{Afonin:2021cwo})
\begin{equation}
\label{b_value}
  b=-\frac12.
\end{equation}

\begin{figure}[!b]
  \begin{center}
    \includegraphics[width=\textwidth]{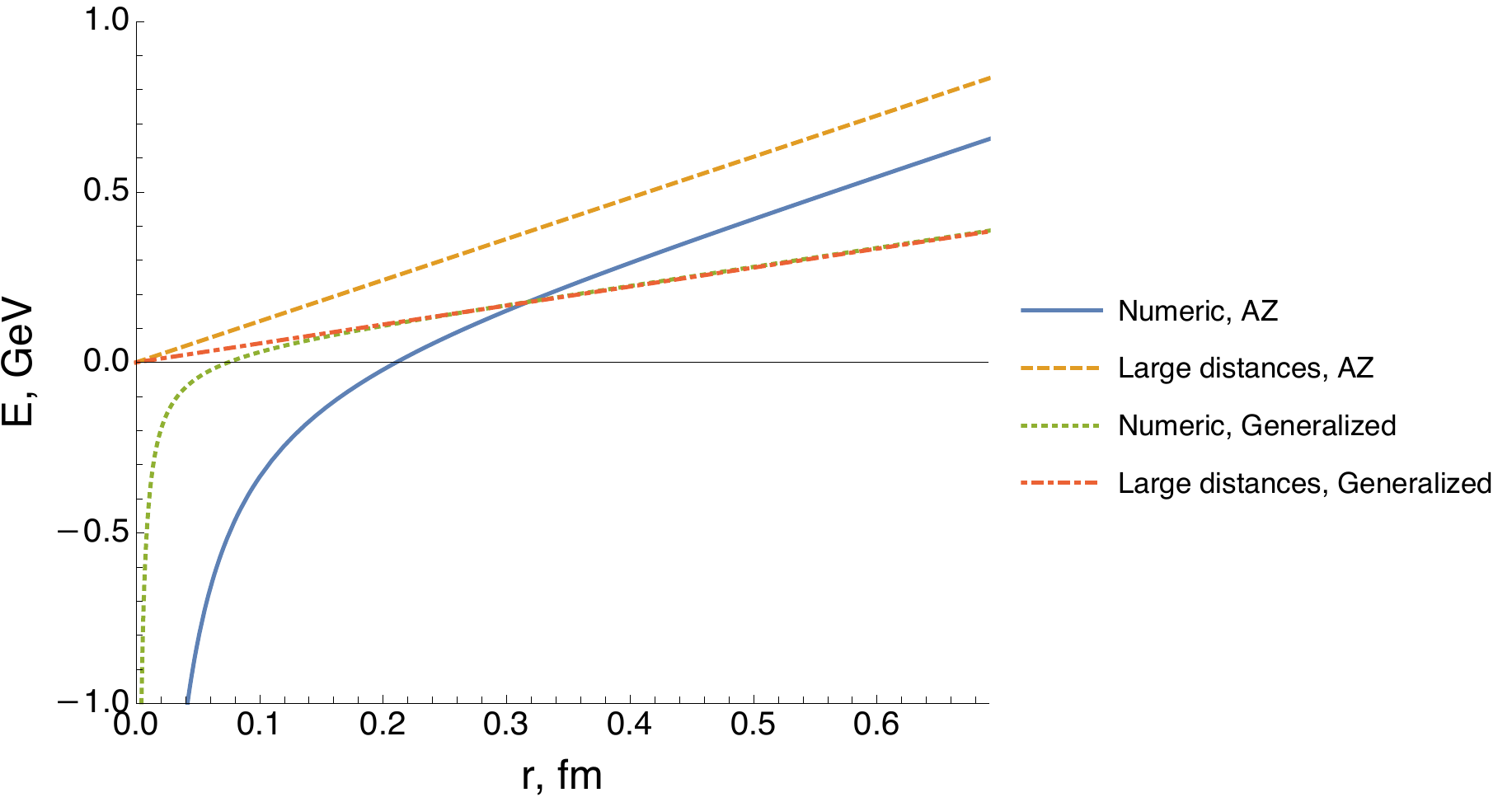}
  \end{center}
  \vspace{-6mm}
  \caption{Comparison between the \textit{large} distance asymptotics of the potential and
  the numerical evaluation of the exact integration formulae (such as~\eqref{r_int_def_2}).
  ``AZ'' stands for~\cite{Andreev:2006ct} with adjustments described in the
  Appendix~\ref{adj_app}, ``Generalized'' --- this work. The parameters are set to:
  \(b=-0.5\), the dilaton parameter to \(4c=1.1\,\text{GeV}^2\), and the normalization
  constant to \(R^2/\alpha'=1\).}
  \label{large_r_comp_plot}
\end{figure}

The parameter $c$ can be fixed from the mean slope of Regge like spectra of light mesons
found in the compilation~\cite{bugg} (see also the review~\cite{Afonin:2007jd}):  \(a\approx1.14\,\text{GeV}^2\).
In our further estimates, we will round this value down to \(a=4c=1.1\,\text{GeV}^2\).
Note in passing that the intercept~\eqref{b_value} in the generalized SW vector spectrum~\eqref{spSW}
leads to the prediction
\begin{equation}
\label{rho}
  m_\rho^2=2c=a/2,
\end{equation}
that correctly reproduces the $\rho$-meson mass.
This is an important phenomenological argument in favor of the physical choice~\eqref{b_value}.
It is interesting to note that the relation~\eqref{rho} is automatically satisfied in the
light-front holographic QCD~\cite{br3} with the same value of slope \(a=1.1\,\text{GeV}^2\)~\cite{{Brodsky:2016yod}}.
There are several other, unrelated to the holographic QCD, theoretical arguments in favor of the physical
choice~\eqref{b_value}, they are discussed in Ref.~\cite{Afonin:2021cwo}.

Having fixed the parameters we are ready to analyze the impact of non-zero intercept
parameter $b$ (which encodes the value of mass gap and likely the effects of the dynamical
chiral symmetry breaking~\cite{Afonin:2021cwo}) on the large distance
asymptotics of the potential in question.
The Figure~\ref{large_r_comp_plot} demonstrates the difference between the large distance
asymptotics of the potential in the model of~\cite{Andreev:2006ct}, corresponding to $b=0$,
and the more physical case corresponding to~\eqref{b_value}.
The adjustments needed to take into account the difference in the dilaton background definition
in our work and in~\cite{Andreev:2006ct} are described in the Appendix~\ref{adj_app}.
In the Figure~\ref{large_r_comp_plot}, we also
compare the asymptotics with full numerical evaluation of the corresponding original
integrals, such as~\eqref{r_int_def_2}.

\begin{wrapfigure}{r}{0.4\textwidth}
  \vspace{-7mm}
  \begin{center}
    \includegraphics[width=0.38\textwidth]{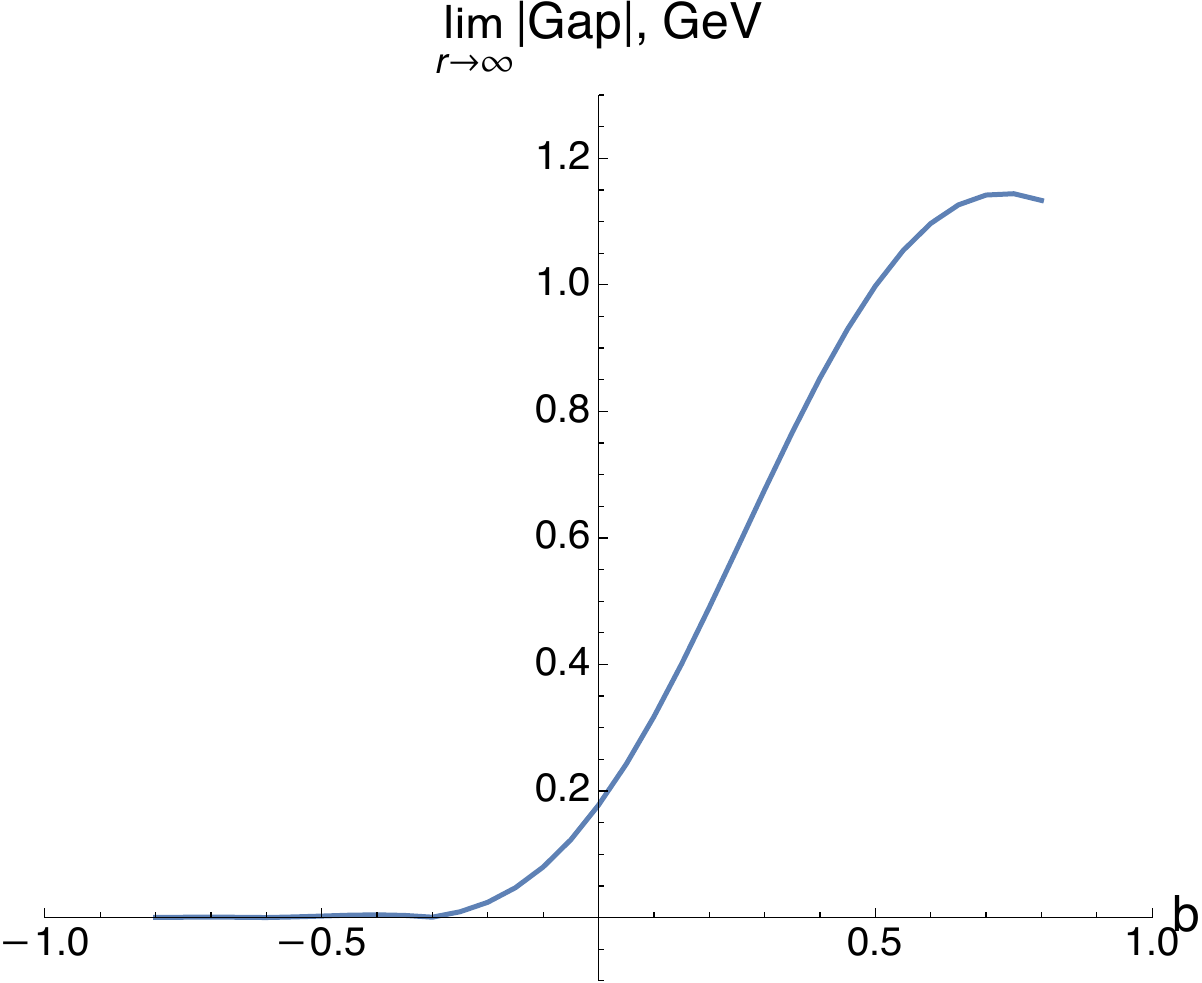}
  \end{center}
  \vspace*{-7mm}
  \caption{The gap limit at \(r\to\infty\).}
  \vspace*{-5mm}
  \label{gap_limit}
\end{wrapfigure}

As follows from the plots~\ref{large_r_comp_plot}, we obtained a gap between the exact numerical and asymptotic
solutions in Andreev--Zakharov's model~\cite{Andreev:2006ct}. It is seen also that
this gap disappears in the physical case of \(b=-0.5\). In the general case, the given gap
turns out to be \(b\)--dependent. We display this dependence in Figure~\ref{gap_limit}.
We could not trace analytically the appearance of this discrepancy as a function of \(b\).

In summary, the Figure~\ref{large_r_comp_plot} shows that the incorporation of negative intercept
parameter leads to a smaller slope (usually associated with the hadron string tension)
of linearly rising confinement potential.

\section{Small distance potential}\label{small_r_sect}

Consider now the asymptotics of the energy $E(r)$ at small distances $r$. Since the distance \(r\)
is the growing function of \(\lambda\) by virtue of exponents, the small distances
correspond to small values of the parameter \(\lambda\). From~\eqref{r_int_def} we can
deduce the small \(\lambda\) asymptotics of \(r\) by first expanding the integrand at
\(\lambda\to0\)
\begin{equation}
  r\underset{\lambda\to0}{=}
  2\sqrt{\frac{\lambda}{c}}\int\displaylimits_{0}^{1}dv\,\lsb
  \frac{v^2}{\sqrt{1-v^4}}+
  2\lambda\lb1+2\frac{U'(b,0,0)}{U(b,0,0)}\rb\frac{v^2(1-v^2)}{(1-v^4)^{3/2}}+
  O(\lambda^2)\rsb.
\end{equation}
Each term in the integral can be integrated separately using the integral representation of the Euler beta-function,
as was done, for example, in Ref.~\cite{Andreev:2006ct}. However, instead we would like to
take a moment to point out a small subtlety with this integration and perform it more
carefully.

First, we should note that the well-known integral representation of the beta-function,
\begin{equation*}
  B(a,b)\equiv\int\limits_0^1dt\,t^{a-1}(1-t)^{b-1},
\end{equation*}
is only valid if both arguments, \(a\) and \(b\), are greater than zero, which makes
this representation inapplicable in our case, since, as we will show shortly, the
arguments can indeed take negative value. Instead, we propose to use the integral
representation of the \textit{reduced} beta-function,
\begin{equation}
  B_x(a,b)\equiv\int\limits_0^xdt\,t^{a-1}(1-t)^{b-1},
\end{equation}
which, in addition, can be defined for negative values of the second argument \(b\) as long as
\(x<1\). We perform the actual integration with the use of the shorthand formula
\begin{equation}\label{red_beta_integral}
  \int\limits_0^1 dv\, v^a (1-v^4)^b=
  \frac{1}{4}\lim_{x\to1}B_x\lb\frac{a+1}{4},b+1\rb,
\end{equation}
which yields
\begin{equation}\label{r_lambda_0_beta}
  r\underset{\lambda\to0}{=}\frac{1}{2}\sqrt{\frac{\lambda}{c}}\lim_{x\to1}\lsb
  B_x\lb\frac{3}{4},\frac{1}{2}\rb+
  \lambda\lb1+2\frac{U'(b,0,0)}{U(b,0,0)}\rb\lb B_x\lb\frac{3}{4},-\frac{1}{2}\rb-
  B_x\lb\frac{5}{4},-\frac{1}{2}\rb\rb\rsb.
\end{equation}
In order to take the limit we make use of the expansion of the reduced beta-function
\begin{equation}\label{red_beta_exp_1}
  B_x(a,b)\underset{x\to1}{=}B(a,b)-\frac{(1-x)^b}{b}+O\lb(x-1)^{b+1}\rb,
\end{equation}
where the beta-function \(B(a,b)\) represents the standard combination of the
gamma-functions,
\begin{equation}
  B(a,b)=\frac{\Gamma(a)\Gamma(b)}{\Gamma(a+b)}.
\end{equation}
It is seen immediately that the first term in~\eqref{r_lambda_0_beta} reduces
to the normal beta-function and, more importantly, the divergences from the second
and the third terms cancel each other out. Note that while in this particular case it
does not make difference, this subtle detail would have been missed if we
had used the normal beta-function instead of the reduced one from the beginning.

In the final step, using the standard properties of the beta- and gamma-functions and
introducing a new parameter
\begin{equation}\label{rho_def}
  \rho\equiv\frac{\Gamma\lb\frac{1}{4}\rb^2}{(2\pi)^{3/2}},
\end{equation}
we obtain the small-\(\lambda\) expansion of the distance
\begin{equation}\label{r_small_lambda}
  r\underset{\lambda\to0}{=}
  \frac{1}{\rho}\sqrt{\frac{\lambda}{c}}\lsb1+
  \lambda\lb1+2\frac{U'(b,0,0)}{U(b,0,0)}\rb\lb\pi\rho^2-1\rb\rsb.
\end{equation}

The small-\(\lambda\) asymptotics of the integral for the energy~\eqref{en_int_def} can
be analyzed in a similar manner. The final result is
\begin{equation}\label{en_small_lambda}
  E\underset{\lambda\to0}{=}
  \frac{R^2}{2\pi\alpha'\rho}U^4(b,0,0)\sqrt{\frac{c}{\lambda}}\lsb
  -1+\lambda\lb1+2\frac{U'(b,0,0)}{U(b,0,0)}\rb\lb3\pi\rho^2-1\rb\rsb.
\end{equation}
The explicit formulae for the intermediate steps of this computation can be found in the
Appendix~\ref{small_r_app}.

In order to obtain the small distance potential we combine the
last two expressions by expressing \(\lambda\) in terms of \(r\). First, we extract
from~\eqref{r_small_lambda}
\begin{equation}\label{sqrt_asymp}
  \sqrt{\frac{c}{\lambda}}=\frac{1}{\rho r}\lsb1+
  \lambda\lb1+2\frac{U'(b,0,0)}{U(b,0,0)}\rb\lb\pi\rho^2-1\rb\rsb,
\end{equation}
which we substitute back into the asymptotics~\eqref{en_small_lambda} for the energy
\begin{multline}
  E\underset{\lambda\to0}{=}
  \frac{R^2}{2\pi\alpha'\rho}\frac{U^4(b,0,0)}{\rho r}\lsb1+
  \lambda\lb1+2\frac{U'(b,0,0)}{U(b,0,0)}\rb\lb\pi\rho^2-1\rb\rsb\times\\\times\lsb
  -1+\lambda\lb1+2\frac{U'(b,0,0)}{U(b,0,0)}\rb\lb3\pi\rho^2-1\rb\rsb.
\end{multline}
Next, we perform the multiplication of the square brackets while retaining only the terms
up to quadratic in \(\lambda\),
\begin{equation}\label{pot_small_r_interm}
  \begin{aligned}
    E&\underset{\lambda\to0}{=}
    \frac{R^2}{2\pi\alpha'\rho}\frac{U^4(b,0,0)}{\rho r}\lsb-1+
    \lambda\lb1+2\frac{U'(b,0,0)}{U(b,0,0)}\rb\lb-\pi\rho^2+1+3\pi\rho^2-1\rb\rsb=\\&=
    \frac{R^2}{2\pi\alpha'\rho}\frac{U^4(b,0,0)}{\rho r}\lsb-1+
    2\pi\rho^2\lambda\lb1+2\frac{U'(b,0,0)}{U(b,0,0)}\rb\rsb.
  \end{aligned}
\end{equation}
Finally, we express from~\eqref{sqrt_asymp} in the leading order
\begin{equation}
  \lambda=c\rho^2r^2,
\end{equation}
substitute this into~\eqref{pot_small_r_interm}, perform some simplifactions, and we get
\begin{equation}
  E\underset{r\to0}{=}\frac{R^2}{\alpha'}U^4(b,0,0)\lsb
  -\frac{1}{2\pi\rho^2}\frac{1}{r}+
  \lb1+2\frac{U'(b,0,0)}{U(b,0,0)}\rb c\rho^2r\rsb.
\end{equation}
This result can be simplified even further by using the properties of the Tricomi
function. First, we note that
\begin{equation}
  U(b,k,0)=\frac{\Gamma(1-k)}{\Gamma(b-k+1)},\quad\text{if }k<1,
\end{equation}
which means that
\begin{equation}
  U^4(b,0,0)=\frac{1}{\Gamma(1+b)^4}.
\end{equation}
Secondly, using various known series expansions of the Tricomi function we can write
\begin{equation}
\label{tric_exp}
  1+2\frac{U'(b,0,0)}{U(b,0,0)}= 1+2b\psi(b+1)+4b\gamma+2b\log\varepsilon,
\end{equation}
where $\varepsilon\rightarrow 0$. Here \(\psi\) is the digamma function and \(\gamma\)
is Euler's constant. The last diverging constant appears from the expansion of
$U'(b,0,\varepsilon)$ at small $\varepsilon$. Combining these expressions we obtain
\begin{equation}
\label{pot_small_r}
  E\underset{r\to0}{=}\frac{R^2}{\alpha'}\lsb-\frac{\kappa_0}{r}+\sigma_0 r\rsb,
\end{equation}
\begin{equation}
\label{small_r_not}
  \kappa_0\equiv\frac{1}{\Gamma(1+b)^4}\frac{1}{2\pi\rho^2},\quad
  \sigma_0\equiv\frac{1+2b\psi(b+1)+4b\gamma}{\Gamma(1+b)^4}c\rho^2.
\end{equation}
Note that in the relation~\eqref{pot_small_r} we subtracted the logarithmic divergence in the expansion~\eqref{tric_exp}
arising from the limit of \(\lambda\to 0\) in our notation. The expression~\eqref{pot_small_r}
represents, thus, the {\it renormalized} energy.

\begin{figure}[!t]
  \begin{center}
    \includegraphics[width=\textwidth]{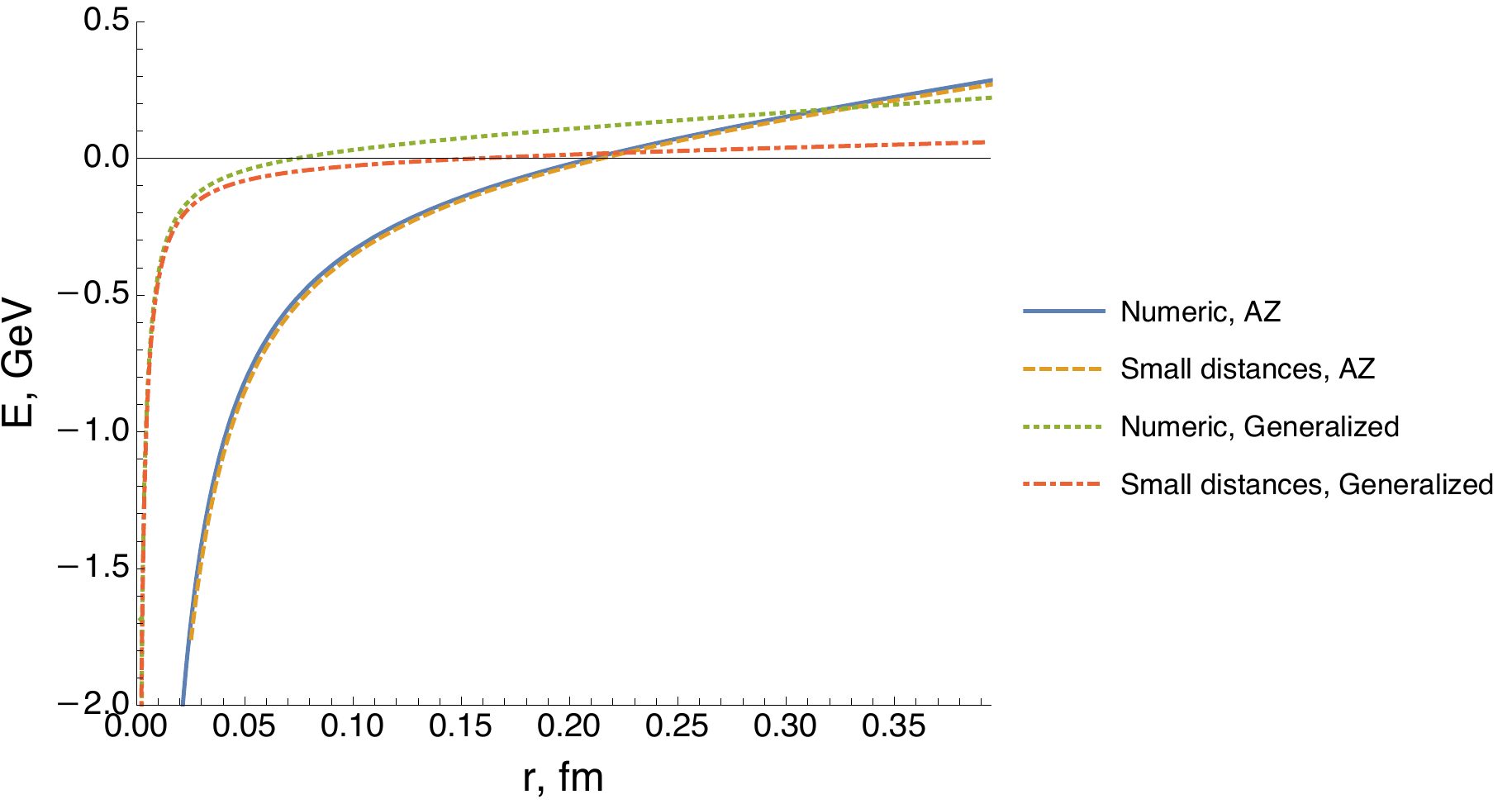}
  \end{center}
  \vspace{-6mm}
  \caption{Comparison between the \textit{small} distance asymptotics of the potential and
  the numerical evaluation of the exact integration formulae (such as~\eqref{r_int_def_2}).
  ``AZ'' stands for~\cite{Andreev:2006ct} with adjustments described in the
  Appendix~\ref{adj_app}, ``Generalized'' --- this work. The parameters are set to:
  \(b=-0.5\), the dilaton parameter to \(4c=1.1\,\text{GeV}^2\), and the normalization
  constant to \(R^2/\alpha'=1\).}
  \label{small_r_comp_plots}
\end{figure}

Let us now compare the small distance potential~\eqref{pot_small_r} with the one
obtained in~\cite{Andreev:2006ct}. As in the previous section, for the latter we use
the adjusted formulae which are presented in the Appendix~\ref{adj_app}. The comparison
is given in the Figure~\ref{small_r_comp_plots}. As a benchmark we also demonstrate
the results of the numerical evaluation of the corresponding exact formulae for
\(r(\lambda)\) and \(E(\lambda)\). One can note that the gap between asymptotics and
numerical results that we observed for large distances for the model of~\cite{Andreev:2006ct}
disappears at small distances. Additionally, the small distance potential asymptotics in
the current model reproduce the ``exact'' solutions quite well.

In summary, the Figure~\ref{small_r_comp_plots} demonstrates that the Coulomb part of the potential comes into play
at smaller distances if the intercept parameter $b$ is negative.
For instance, if we set the normalization constant to \(R^2/\alpha'=1\) as in the Figure~\ref{small_r_comp_plots}
then in the Andreev--Zakharov's model, defined at $b=0$, the Coulomb part becomes substantial at a distance less than 0.1~fm
while in the considered phenomenologically preferable case of \(b=-0.5\),
this distance drops to about 0.03~fm.

\section{The scalar case}

Let us now apply the analysis above to the scalar case.
In this case, the analogue of the generalized SW action~\eqref{gen_sw} reads~\cite{Afonin:2021cwo}
\begin{equation}
\label{gen_sw_s}
  S_\text{sc}=\int d^4xdz\sqrt{g}e^{-cz^2}U^2(b,-1,|cz^2|)\mathcal{L}_\text{sc}.
\end{equation}
The mass spectrum of the SW model~\eqref{gen_sw_s} with $c>0$ is~\cite{Afonin:2021cwo,afonin2020}
\begin{equation}
\label{spSW_sc}
m_n^2=4c(n+\Delta/2+b),\qquad n=0,1,2,\dots,
\end{equation}
where $\Delta$ is the canonical dimension of QCD scalar operator dual to the corresponding 5D scalar field in the
Lagrangian density $\mathcal{L}_\text{sc}$.

As in the vector case, for our purposes one should rewrite the dilaton background as certain modification of AdS$_5$ metric
using~\eqref{tr} and continue the modified metric to the Euclidean space. The background
function \(h(z)\) of the metric in~\eqref{bg_f} is then replaced by (see also discussions in~\cite{Afonin:2021zdu})
\begin{equation}
\label{bg_f_s}
  h=e^{2cz^2/3}U^{4/3}(b,-1,cz^2)=e^{2\lambda v^2/3}U^{4/3}(b,-1,\lambda v^2).
\end{equation}
This changes the expressions for \(r(\lambda)\) and \(E(\lambda)\)  to the following ones,
\begin{equation}
\label{r_int_scalar}
  r=2\sqrt{\frac{\lambda}{c}}\int\limits_0^1dv\,
  \frac{U^{4/3}(b,-1,\lambda)}{U^{4/3}(b,-1,\lambda v^2)}
  \frac{v^2e^{2\lambda(1-v^2)/3}}
  {\sqrt{1-v^4e^{4\lambda(1-v^2)/3}\frac{U^{8/3}(b,-1,\lambda)}{U^{8/3}(b,-1,\lambda v^2)}}},
\end{equation}
\begin{equation}\label{en_int_scalar}
  E=\frac{R^2}{\pi\alpha'}\sqrt{\frac{c}{\lambda}}\lsb\int\limits_0^1\frac{dv}{v^2}\lb
  \frac{e^{2\lambda v^2/3}U^{4/3}(b,-1,\lambda v^2)}{\sqrt{1-v^4e^{4\lambda(1-v^2)/3}\frac{U^{8/3}(b,-1,\lambda)}{U^{8/3}(b,-1,\lambda v^2)}}}-D\rb-D\rsb,
\end{equation}
with the regularization constant \(D\equiv U^{4/3}(b,-1,0)\).

The derivation of the asymptotics in question is essentially the same.
For the large distance asymptotics we get
\begin{equation}
  E\underset{r\to\infty}{=}\frac{R^2}{\alpha'}\sigma_\infty r,
\end{equation}
\begin{equation}
\label{large_r_not_s}
  \sigma_\infty=\frac{e^{2x/3}U^{4/3}(b,-1,x)}{2\pi x}c,
\end{equation}
where \(x\) (see~\eqref{eq12b}) is now the solution to the equation
\begin{equation}\label{realscalar}
  1-\frac{2}{3}x-\frac{4}{3}x\frac{U'(b,-1,x)}{U(b,-1,x)}=0,
\end{equation}
which is the scalar counterpart of Eq.~\eqref{eq12}.
In the case of the standard SW model, $b=0$, we get $x(0)=3/2$.
Comparing~\eqref{large_r_not_s} with~\eqref{large_r_not} at $b=0$ we
have $\sigma_\infty^\text{(sc)}/\sigma_\infty^\text{(vec)}=1/3$.
The emerging factor of $1/3$ stems from the appearance of the additional
factor of $1/3$ in the exponent of scalar background function~\eqref{bg_f_s}
in comparison with~\eqref{bg_f}. The given factor provides the equal slope of
scalar and vector radial Regge trajectories in the SW model.

The details of the derivation of the small distance asymptotics are given in the
Appendix~\ref{small_r_app}, the result is
\begin{equation}
\label{en_small_lambda_scalar}
  E\underset{r\to0}{=}\frac{R^2}{\alpha'}U^{4/3}(b,-1,0)\lsb
  -\frac{1}{2\pi\rho^2}\frac{1}{r}+
  \lb1+2\frac{U'(b,-1,0)}{U(b,-1,0)}\rb\frac{c\rho^2}{3}r\rsb.
\end{equation}
It can be simplified using the analogous to~\eqref{tric_exp} expansion,
\begin{equation}
\label{tric_exp_scalar}
  1+2\frac{U'(b,-1,0)}{U(b,-1,0)}= 1-2b,
\end{equation}
where in contrast to the vector case the logarithmic term has the form of \(\varepsilon \log\varepsilon\)
that goes to \(0\) as \(\varepsilon\rightarrow 0\).
Finally, we obtain
\begin{equation}
  E\underset{r\to0}{=}\frac{R^2}{\alpha'}\lsb-\frac{\kappa_0}{r}+\sigma_0 r\rsb,
\end{equation}
\begin{equation}
  \kappa_0=\frac{1}{\Gamma(2+b)^{4/3}}\frac{1}{2\pi\rho^2},\quad
  \sigma_0=\frac{1-2b}{\Gamma(2+b)^{4/3}}\frac{c\rho^2}{3}.
\end{equation}
It is interesting to note that in the scalar case, the logarithmic divergence from the
derivative of the Tricomi in~\eqref{tric_exp_scalar}
is absent, i.e., the asymptotics~\eqref{en_small_lambda_scalar}
does not need the renormalization caused by the subtraction of that divergence.

\section{Discussions}

For a phenomenological analysis of our results we should first provide
a very brief review of the relevant phenomenology of the Cornell potential~\eqref{cornell}. Let us
write this potential once more for a further convenience,
\begin{equation}
\label{cornell2}
  V(r)=-\frac{\kappa}{r}+\sigma r+C.
\end{equation}
In typical potential models for heavy quarkonia (for a review see, e.g.,~\cite{bali}),
the constant $C$ is roughly $C\approx-0.3$~GeV.
The potential~\eqref{cornell2} was first proposed in Ref.~\cite{Eichten:1978tg} for a non-relativistic
description of the charmonia spectrum. The linear confinement potential at large distances was inferred from lattice gauge theory
and also inspired by the dual string model~\cite{dual1,dual2,dual3,dual4}.
The spin averaged charmonia spectrum (plus ground states of bottomonia)
results in parameter values~\cite{bali}
\begin{equation}
\label{charm}
\text{Charmonia:}\qquad  \kappa\approx0.25,\quad
  \sigma\approx0.21\,\text{GeV}^2.
\end{equation}
The perturbative QCD predicts the value of Coulomb coefficients for quarkonia,
\begin{equation}
\label{run}
\kappa=\frac{4}{3}\alpha_s(r),
\end{equation}
where $\alpha_s(r)$ is the QCD running coupling which depends on renormalization scheme.
The renormalization scheme to be used with the potential is the so-called V-scheme which is determined by
perturbative high order QCD corrections to the static potential (see, e.g., Refs.~\cite{Deur:2016tte,Kataev:2015yha} for relevant  discussions).
In the case of Cornell potential, however, the scale for $\alpha_s(r)$ is not well-defined and
the coefficient $\kappa$ of the Coulomb-like part is related with an average value $\langle\alpha_s\rangle$
over the scale range where that part of the potential is dominant. The mean value of the coupling $\langle\alpha_s\rangle$
depends on the size of the hadrons considered.
The value of $\langle\alpha_s\rangle$ extracted in Ref.~\cite{Eichten:1978tg} from hadronic decays of excited charmonia is
$\langle\alpha_s(\psi)\rangle=0.19\pm0.03$. This value is consistent with~\eqref{charm} and~\eqref{run}: $\frac43 0.19\approx0.25$.
After inclusion of excited bottomonia, that probe the potential at smaller distances, the estimates~\eqref{charm}
were shifted to the following approximate values~\cite{bali},
\begin{equation}
\label{cb}
\text{Charmonia}+\text{Bottomonia:}\qquad  \kappa\approx0.51,\quad
  \sigma\approx0.18\,\text{GeV}^2.
\end{equation}
In the region $0.2\,\text{fm}<r<1\,\text{fm}$, which is effectively probed by spin-averaged quarkonia splittings,
the parametrizations~\eqref{charm} and~\eqref{cb}, however, differ only marginally: The higher value of the
Coulomb coefficient is compensated for by a smaller slope $\sigma$~\cite{bali}.

The value of slope \(\sigma\) in~\eqref{cb} turns out to be remarkably insensitive to a chosen distance.
The same value is typically used for the description of the light meson spectrum within the potential models\footnote{With certain
model modifications like a smearing procedure in the coordinate space to take into account relativistic effects and
the use of an {\it ad hoc} model for the running coupling that freezes out at low energies, the paper~\cite{Godfrey:1985xj}
is a classical work in this field.}. Even more surprising is the quantitative agreement of the mean slope of Regge spectra of light mesons,
\(a\approx1.14\,\text{GeV}^2\)~\cite{bugg}, with the slope predicted by semiclassical quantization of hadron string with
linearly rising potential energy~\eqref{cornell2}, \(a=2\pi\sigma\), where \(\sigma\) is the string tension (see, e.g., the discussions in Ref.~\cite{Afonin:2007jd}),
if the value from~\eqref{cb} is used for \(\sigma\).

After this brief reminder, we are ready to analyze the relevant phenomenological predictions of our
generalized SW holographic model. First of all, it should be noted that the coupling of the Coulomb-like part of the potential obtained within
the holographic approach is not running at large energy-momenta because this approach represents a classical framework.
On the other hand, it can run at the low energy-momenta due to semiclassical effects that are part of the definition of the coupling~\cite{Brodsky:2010ur}.
Therefore, the constant coupling should be understood as the value averaged over energy-momentum of the actually running coupling.
With this in mind, comparing directly our results to that of the Cornell potential is self-consistent.

\begin{figure}[tb]
  \begin{center}
     \includegraphics[width=\textwidth]{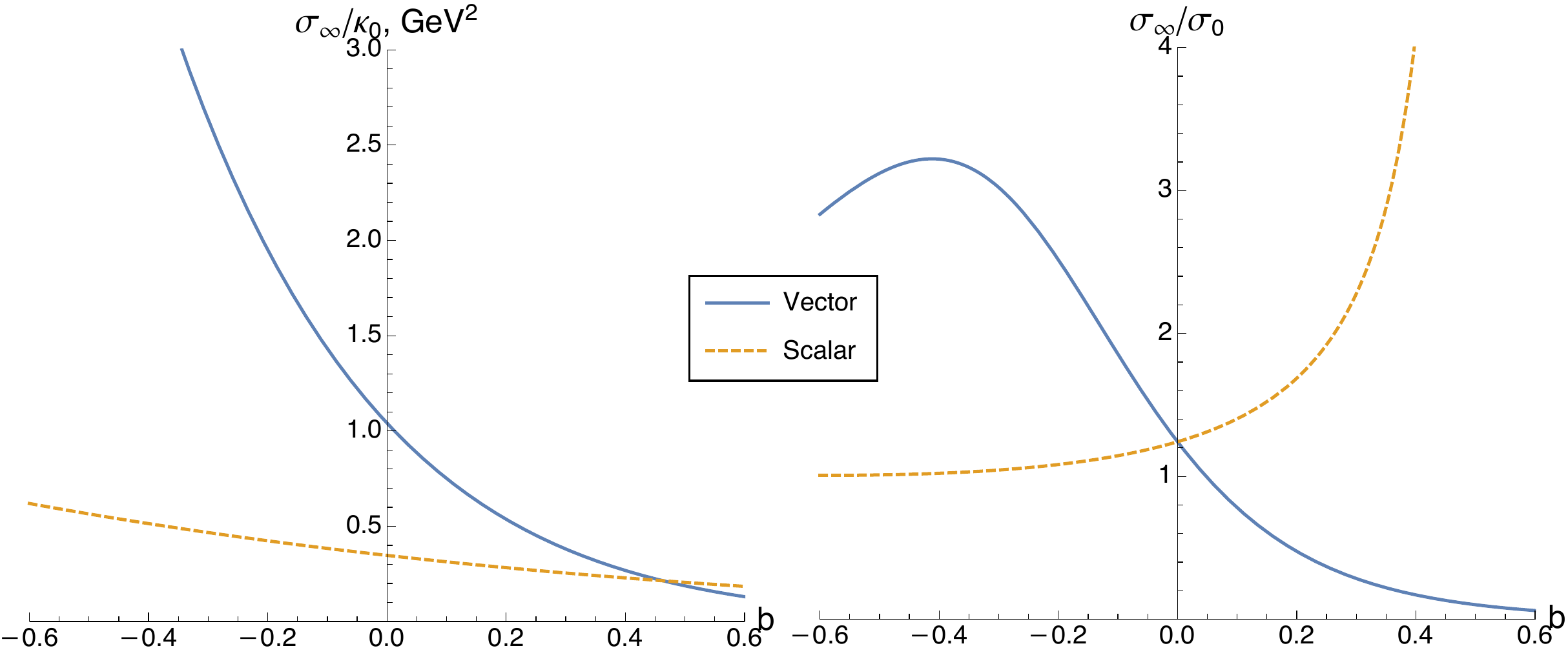}
  \end{center}
  \vspace{-6mm}
  \caption{The ratios $\sigma_\infty/\kappa_0$ (left) and $\sigma_\infty/\sigma_0$ (right) in the
  vector and scalar cases. The Regge slope is set to \(4c=1.1\,\text{GeV}^2\).}
  \label{coeff_ratios_plots}
\end{figure}

Let us consider the predictions which do not depend on the general
normalization constant \(R^2/\alpha'\). We can temporary set \(R^2/\alpha'=1\), then our definitions
of parameters $\kappa$ and $\sigma$ coincide with the definitions in the Cornell potential~\eqref{cornell2}.
The first prediction is the ratio $\sigma_\infty/\kappa_0$. In the vector case, we obtain from~\eqref{large_r_not}
and~\eqref{small_r_not}
\begin{equation}
\label{sigma_kappa_0}
  \frac{\sigma_\infty}{\kappa_0}=\frac{\Gamma\lb\frac{1}{4}\rb^4\Gamma(1+b)^4e^{2x}U(b,0,x)^4c}{(2\pi)^3 x}.
\end{equation}
Andreev and Zakharov got the estimate $\sigma_\infty/\kappa_0\approx0.85\,\text{GeV}^2$~\cite{Andreev:2006ct}
and found it ``disappointing'' that this prediction disagreed significantly with the corresponding phenomenological
ratio from~\eqref{cb},
\begin{equation}
\label{cb2}
\text{Charmonia}+\text{Bottomonia:}\qquad  \frac{\sigma}{\kappa}\approx0.35\,\text{GeV}^2.
\end{equation}
We should note, however, that in the fit~\eqref{charm} this ratio is equal to $0.84$ in
nice agreement with the Andreev--Zakharov prediction. Taking into account the aforementioned remark on
a phenomenological proximity of the parametrizations~\eqref{charm} and~\eqref{cb}, the obtained estimate looks
reasonable\footnote{We should mention, however, that the value of Regge slope was set in~\cite{Andreev:2006ct}
equal to $a=4c=0.9\,\text{GeV}^2$. With our value of \(4c=1.1\,\text{GeV}^2\), the relation~\eqref{sigma_kappa_0}
for $b=0$ would give $\sigma_\infty/\kappa_0\approx1.04\,\text{GeV}^2$, see also the Figure~\ref{coeff_ratios_plots}.}.

The general behavior of ratio $\sigma_\infty/\kappa_0$ as a function of intercept parameter $b$
in the
vector and scalar case is displayed in the Figure~\ref{coeff_ratios_plots}. It is seen that the
physical value of intercept in the vector case, $b\approx-0.5$, leads to an unrealistically large
ratio, while the scalar case with $b\approx0$ is perfectly compatible with the phenomenological output~\eqref{cb2}.

The next important ratio is $\sigma_\infty/\sigma_0$ which does not depend on the Regge slope $a=4c$.
The lattice data suggest that the linearly rising part in the Cornell potential~\eqref{cornell2}
is almost universal at large and small distances, hence, a reasonable prediction for $\sigma_\infty/\sigma_0$
should lie near 1~\cite{Andreev:2006ct}. The prediction of Andreev--Zakharov model
is $\sigma_\infty/\sigma_0\approx1.24$~\cite{Andreev:2006ct} that corresponds to $b=0$ in our model.
The general behavior of ratio $\sigma_\infty/\sigma_0$ as a function of $b$ is displayed in
the Figure~\ref{coeff_ratios_plots}. In the same Figure, we show the corresponding behavior in
the scalar SW model. It is seen that the
physical value of intercept in the vector case, $b\approx-0.5$, leads to an unrealistically large
ratio, while the scalar case gives a stable prediction near 1 for $b\lesssim0$.

\begin{figure}[!t]
  \begin{center}
    \includegraphics[width=1\textwidth]{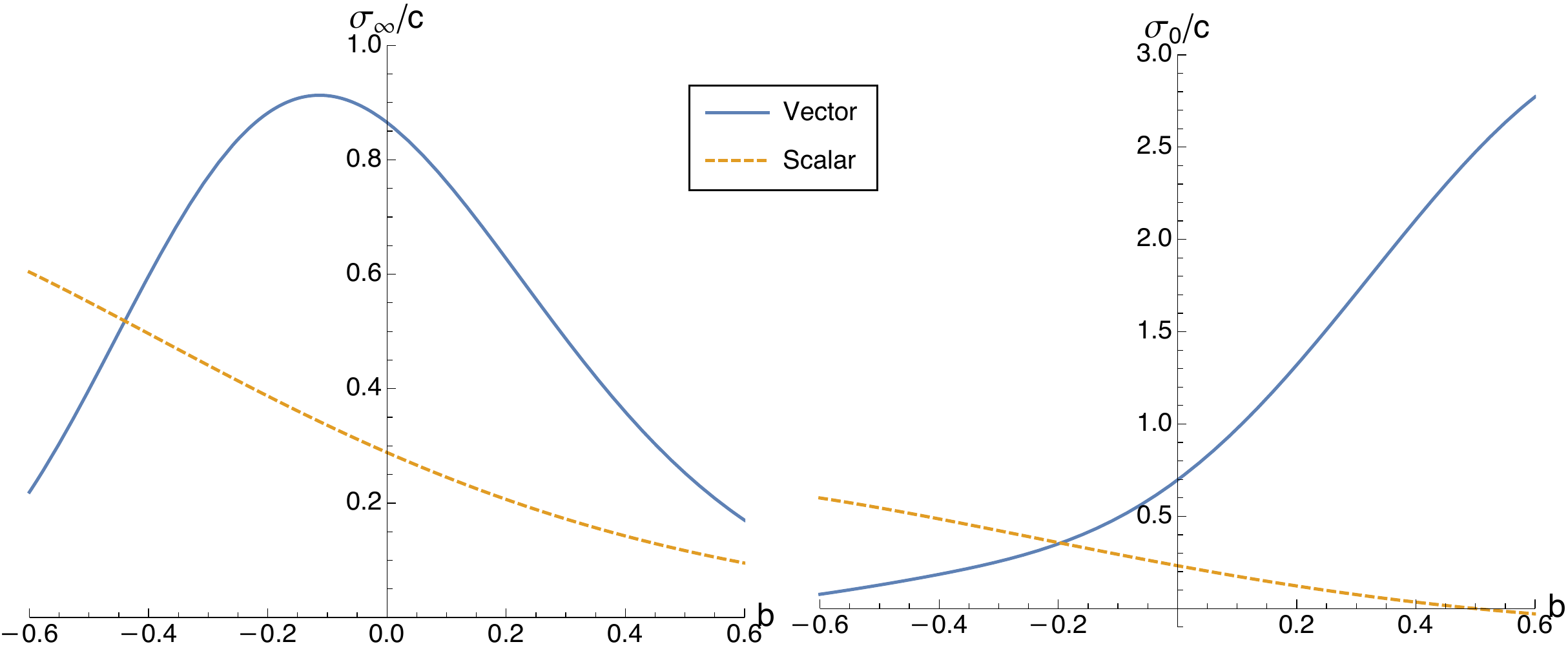}
  \end{center}
  \vspace{-6mm}
  \caption{The ratios $\sigma_\infty/c$  (left) and $\sigma_0/c$ (right) in the vector and scalar cases.}
  \label{coeff_ratios_plots3}
\end{figure}

Our further predictions will depend on the choice of
normalization constant \(R^2/\alpha'\). First let us assume that this constant is $b$--independent.
Without loss of generality, we may again set \(R^2/\alpha'=1\). The slope $\sigma$ is a function
of $b$. Since the intercept $b$ encodes important physics of non-perturbative strong interactions,
it is interesting to check the behavior of $\sigma(b)$ at large and small distances since this
reflects the dependence on $b$ of string tension and of the slope of the Regge spectrum.
In the Figure~\ref{coeff_ratios_plots3}, we show the dependence on $b$ of $c$--independent
ratios $\sigma_\infty/c$ and $\sigma_0/c$ in the vector and scalar SW models. One can conclude
from the presented plots that the behavior of $\sigma_\infty(b)$ and $\sigma_0(b)$ is
qualitatively similar in the scalar case and is very different in the vector one.

\begin{figure}[b!]
  \begin{center}
    \includegraphics[width=\textwidth]{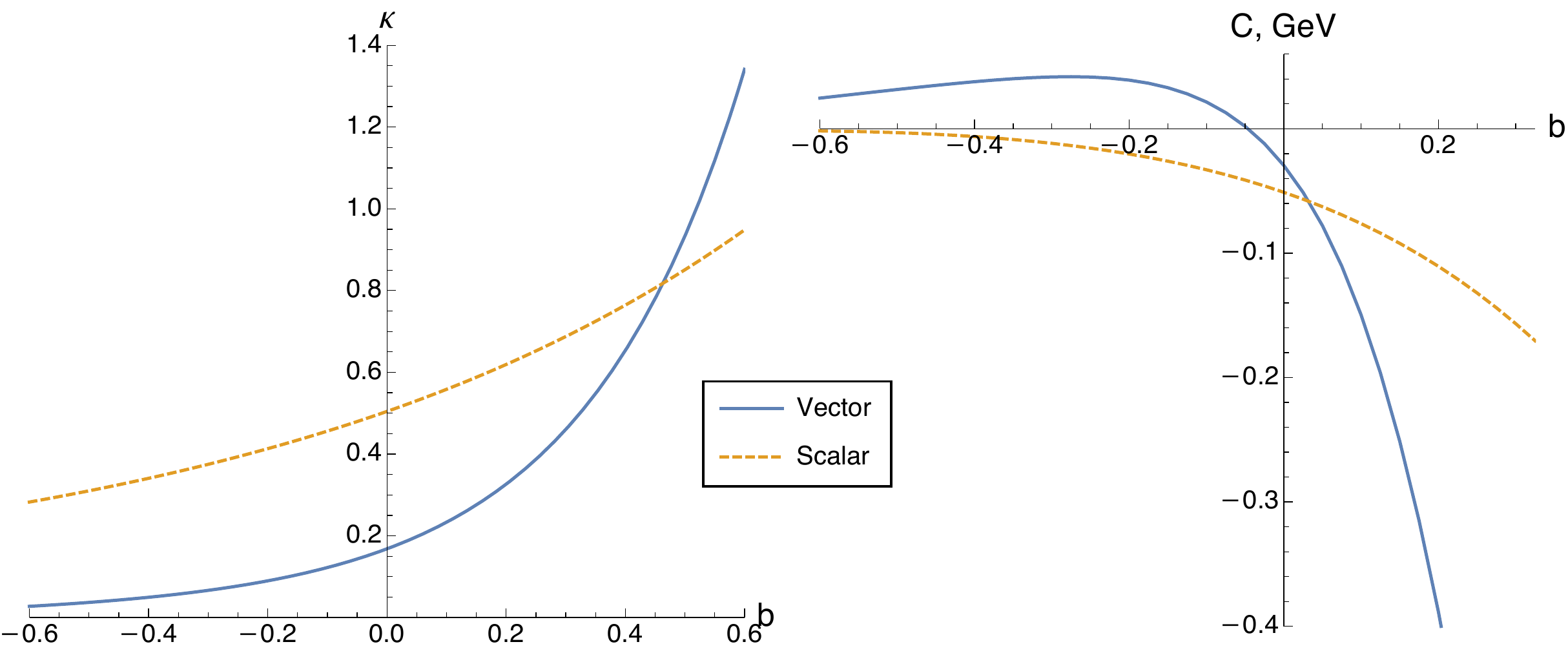}
  \end{center}
  \vspace{-6mm}
  \caption{The behavior of the parameters $\kappa$ (left) and $C$ (right) in the potential~\eqref{cornell2} as functions of the intercept $b$
  when the normalization~\eqref{cons} is imposed. The Regge slope is set to \(4c=1.1\,\text{GeV}^2\) (equivalently, \(\sigma_\infty=0.18\,\text{GeV}^2\)).}
  \label{coeff_ratios_plots4}
\end{figure}

Now let us relax the assumption above. The dependence of the normalization constant \(R^2/\alpha'\) on $b$
can help to amend the following theoretical discrepancy. As was mentioned above, the semiclassical quantization
of hadron string of the Nambu-Goto type with tension $\sigma$ and linearly growing with distance energy as in~\eqref{cornell2},
leads to the slope of the angular and radial Regge mass spectrum \(a=2\pi\sigma\). On the other hand, the same slope
in the SW holographic model is $a=4c$, see, e.g., the SW spectrum~\eqref{8} or~\eqref{spSW_sc}. The physical values of \(\sigma\)
and $c$ remarkably agree with each other. But this agreement can be destroyed by the dependence \(\sigma(b)\) as long as the
parameter $c$ in the SW model does not depend on $b$. And here the normalization constant \(R^2/\alpha'\) can restore
the consistency: The actual $\sigma$ in~\eqref{cornell2} is our $\frac{R^2}{\alpha'}\sigma_\infty$, hence,
we may impose the consistency condition,
\begin{equation}
\label{cons}
\frac{R^2}{\alpha'}2\pi\sigma_\infty(b)=4c,
\end{equation}
that should fix the normalization constant \(R^2/\alpha'\).
For instance, in the vector SW model at $b=0$, we get from~\eqref{large_r_not_0}
and~\eqref{cons} the normalization \(R^2/\alpha'=2/e\approx0.74\).
The given normalization constant is not very distinct from  \(R^2/\alpha'=1\) used
in our work.

When the normalization~\eqref{cons} is imposed, the string tension $\sigma$ in the confinement potential~\eqref{cornell2}
does not depend on the intercept parameter $b$ while the Coulomb parameter $\kappa$ and constant $C$ are functions of $b$.
Their $b$-dependence can be calculated numerically, we display the result in the Figure~\ref{coeff_ratios_plots4}.
These plots show that the physical value of intercept in the vector case, $b\approx-0.5$, leads to an unrealistically small value
of $\kappa$ and to positive $C$, while the scalar case with $b\approx0$ reproduces numerically the value of $\kappa$ in the
phenomenological fit~\eqref{cb} and qualitatively gives the correct sign of constant $C$.

\begin{figure}[!b]
  \begin{center}
    \includegraphics[width=0.6\textwidth]{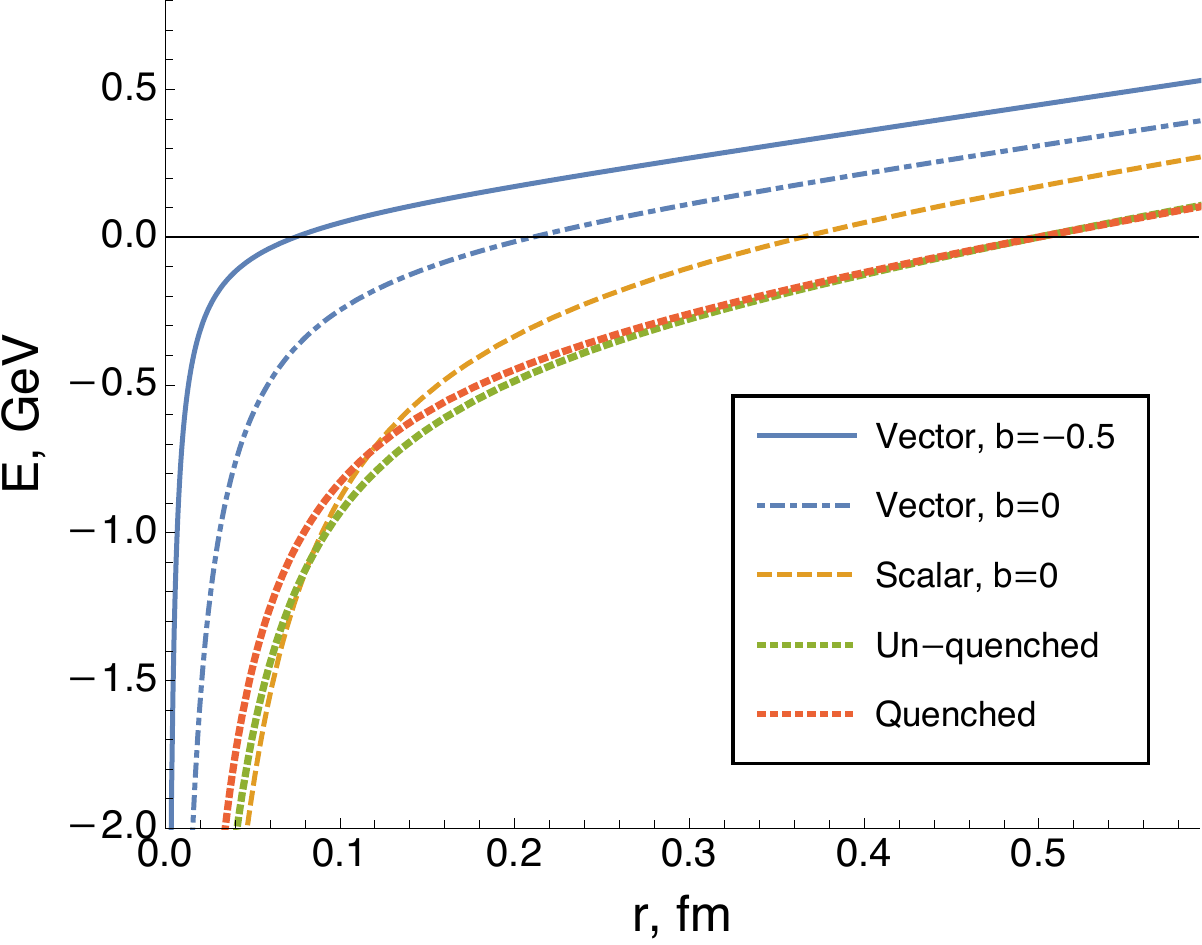}
  \end{center}
  \vspace{-6mm}
  \caption{The behavior of potential energy with distance for three examples of SW holographic model discussed in the text.
  The lattice data for the Wilson action $SU(3)$ potential are taken from the review~\cite{bali}. More precisely, we made
  use of the Cornell potential~\eqref{cornell2} that almost perfectly interpolates these data, the parameters of this potential are the following~\cite{bali}:
  $\sigma=0.18\,\text{GeV}^2$, $\kappa=0.295$ for the quenched approximation and $\kappa\approx0.36$ in the un-quenched case (sea quarks effects are taken into
  account), in both cases the constant $C$ is fixed by the condition
  $E(0.5~\text{fm})=0$.}
  \label{coeff_ratios_plots5}
\end{figure}

It is interesting to compare our models normalized by the condition~\eqref{cons} with the results of lattice simulations in $SU(3)$ gauge theory.
In the Figure~\ref{coeff_ratios_plots5}, we provide such a comparison for three typical cases: The vector SW model with
$b=0$ (the simplest standard variant), with $b=-0.5$ (the phenomenologically preferable variant~\eqref{b_value}), and the scalar SW model with
$b=0$ (the most consistent variant according to our analysis).

In summary, the totality of observations made above shows that the holographic confinement potential quantitatively compatible with the phenomenology
arises in the {\it scalar} version of the SW holographic model, in which the physical value of intercept parameter $b$ in the Regge-like
spectrum~\eqref{spSW_sc} seems to be close to zero.

The latter observation could be converted into a prediction for the mass $m_s$ of ground scalar state.
If the quark physical degrees of freedom are quenched, only the gluon ones are operative.
The minimal dimension of scalar operator constructed from the gluon fields is $\Delta=4$
(the scale-invariant operator $\beta G_{\mu\nu}^2$). Then the spectrum~\eqref{spSW_sc}
with $b\approx0$ combined with the relation~\eqref{rho}
leads to the following prediction for the mass of the lightest scalar glueball,
\begin{equation}
m_s\approx2m_\rho.
\end{equation}
This prediction is close to the mass of the scalar meson $f_0(1500)$ which
is indeed a candidate for the lightest scalar glueball~\cite{pdg}.

\section{Conclusions}

The Soft Wall AdS/QCD approach provides a natural framework for the appearance of the Cornell type confinement potential
at short and long distances. We performed a detailed analytical study of arising confinement potentials within the generalized
version of Soft Wall holographic model in the vector and scalar cases, where the term ``generalized'' means the
incorporation of arbitrary intercept into the radial Regge spectrum.
The intercept parameter is indispensable for a quantitative phenomenological description of the experimental spectrum of light meson resonances.
Our numerical analysis and comparison with the phenomenological Cornell potentials showed that
quantitatively correct confinement potential arises in a consistent way within the scalar Soft Wall holographic model,
while the standard vector version of the model results in a qualitative agreement only.
This conclusion agrees with the recent analysis of Ref.~\cite{Afonin:2018era} where it was argued that
quantitatively correct prediction of the deconfinement temperature within the framework of the same model emerges
in the scalar case. And as in Ref.~\cite{Afonin:2018era}, the found numerical parameters seem to be consistent with the
interpretation of the scalar state $f_0(1500)$~\cite{pdg} as the lightest glueball.

We have no clear understanding on why the scalar version of considered generalized SW holographic model is quantitatively more consistent
with the phenomenology of quark confinement than the vector one. But we see here the manifestation of a general tendency
for the dominance of the vacuum scalar sector in the non-perturbative dynamics of strong interactions. This dominance is ubiquitous:
The universality of hadron-hadron
scattering at ultrahigh energies is believed to be associated with the exchange of scalar pomerons, and not vector gluons, as one might
naively expect; in the opposite limit of very low energies, the predominant part of attraction potential between nucleons is due to the
exchange of scalar $\sigma$-meson~\cite{pdg} (a correlated two-pion exchange), and not pseudoscalar $\pi$-mesons, as one could naively
expect; at intermediate energies, the number of observed scalar isoscalar resonances~\cite{pdg} is much greater than it should follow
from the quark model...

In summary, we believe that our results provide a new demonstration of the key role of the vacuum scalar sector in the description of
confinement physics in strong interactions.

\section*{Acknowledgements}

This research was funded by the Russian Science Foundation grant number 21-12-00020.


\bigskip
\bigskip
\bigskip

\begin{center}
\LARGE\bf Appendices
\end{center}

\appendix

\section{Details of the small distance asymptotics calculation}\label{small_r_app}

In this Appendix, we present technical details concerning the derivation of the small
distance asymptotics.

First we consider the small-\(\lambda\) asymptotics of the energy
integral~\eqref{en_int_def} for the vector case. The expansion of its
integrand at \(\lambda\to 0\), where we also substitute the regularization constant
\(D=U^4(b,0,0)\), reads
\begin{equation}\label{en_small_lambda_exp}
  E\underset{\lambda\to0}{=}
  \frac{R^2}{\pi\alpha'}\sqrt{\frac{c}{\lambda}}\lsb\int\displaylimits_{0}^{1}
  \frac{dv}{v^2}\lb E_0+\lambda E_1+O(\lambda^2)\rb-U^4(b,0,0)\rsb.
\end{equation}
Here we introduced the expansion coefficients
\begin{equation}
  E_0\equiv U^4(b,0,0)\lb\frac{1}{\sqrt{1-v^4}}-1\rb,
\end{equation}
\begin{equation}
  E_1\equiv 2U^4(b,0,0)\lb1+2\frac{U'(b,0,0)}{U(b,0,0)}\rb
  \frac{v^2(1-2v^4+v^2)}{(1-v^4)^{3/2}}.
\end{equation}
On the next step, the \(E_0\) and \(E_1\) terms are integrated using the integral representation
of the reduced beta-function~\eqref{red_beta_integral}. Note, however, that technically
\(E_0\) contains two integrals and both of them are divergent. Below we show a trick which
we use to demonstrate that these divergences cancel each other out. We start by rewriting
the integrals as
\begin{equation}
  \int\displaylimits_{0}^{1}\frac{dv}{v^2}\lb\frac{1}{\sqrt{1-v^4}}-1\rb=
  \lim_{\substack{x\to1\\y\to0}}\int\displaylimits_{y}^{x}\frac{dv}{v^2}\lb\frac{1}{\sqrt{1-v^4}}-1\rb=\dots
\end{equation}
After that we change the integration variable to \(u\equiv v^4\) and split the first integral
into two
\begin{equation}
  \dots=\frac{1}{4}\lim_{\substack{x\to1\\y\to0}}\lb
  \int\displaylimits_0^{x^4}du\frac{u^{-5/4}}{\sqrt{1-u}}-
  \int\displaylimits_0^{y^4}du\frac{u^{-5/4}}{\sqrt{1-u}}-
  \int\displaylimits_{y^4}^{x^4}\frac{du}{u^{5/4}}\rb=\dots
\end{equation}
We can now safely perform the integration (for the first two integrals we use the integral
representation~\eqref{red_beta_integral} of the reduced beta-function) and we get
\begin{equation}
  \dots=\frac{1}{4}\lim_{\substack{x\to1\\y\to0}}\lsb
  B_{x^4}\lb-\frac{1}{4},\frac{1}{2}\rb-
  B_{y^4}\lb-\frac{1}{4},\frac{1}{2}\rb+
  \frac{4}{x}-\frac{4}{y}\rsb=\dots
\end{equation}
Using the expansions of the reduced beta-function~\eqref{red_beta_exp_1} and
\begin{equation}
  B_y(a,b)\underset{y\to0}{=}\frac{y^a}{a}+O(y^{a+1}),
\end{equation}
we obtain
\begin{equation}
  \dots=\frac{1}{4}\lim_{\substack{x\to1\\y\to0}}\lsb
  B\lb-\frac{1}{4},\frac{1}{2}\rb-2\sqrt{1-x^4}+O\lb(1-x^4)^{3/2}\rb+
  \frac{4}{y}+O\lb y^3\rb+
  \frac{4}{x}-\frac{4}{y}\rsb=\dots,
\end{equation}
from which we immediately see that the divergences indeed cancel each other out and the
final result of the integration of \(E_0\) is
\begin{equation}\label{E_0}
  \int\displaylimits_{0}^{1}\frac{dv}{v^2}\,E_0=
  U^4(b,0,0)\lsb\frac{1}{4}B\lb-\frac{1}{4},\frac{1}{2}\rb+1\rsb.
\end{equation}
The integral for \(E_1\) can be represented as three separate integrals,
\begin{equation}
  \int\displaylimits_{0}^{1}\frac{dv}{v^2}\,E_1=
  2U^4(b,0,0)\lb1+2\frac{U'(b,0,0)}{U(b,0,0)}\rb
  \int\displaylimits_{0}^{1}dv\,\frac{1-2v^4+v^2}{(1-v^4)^{3/2}},
\end{equation}
each of which results in the beta-function. The end result for \(E_1\) is
\begin{equation}\label{E_1}
  \int\displaylimits_{0}^{1}\frac{dv}{v^2}E_1=
  \frac{U^4(b,0,0)}{2}\lb1+2\frac{U'(b,0,0)}{U(b,0,0)}\rb\lsb
  B\lb\frac{1}{4},-\frac{1}{2}\rb+B\lb\frac{3}{4},-\frac{1}{2}\rb-
  2B\lb\frac{5}{4},-\frac{1}{2}\rb\rsb.
\end{equation}
We can now substitute~\eqref{E_0} and \eqref{E_1} back into~\eqref{en_small_lambda_exp}
(note that the out-of-integral \(U^4(b,0,0)\) terms cancel each other out) and use the
properties of the gamma-functions as well as the definition~\eqref{rho_def} of \(\rho\).
This leads to the final form~\eqref{en_small_lambda} for the small-\(\lambda\) asymptotics of energy.

The calculations in the scalar case are similar --- we have to adjust only the coefficients in front of the integrals,
which give rise to the beta-functions. The small \(\lambda\) expansion of the
integral~\eqref{r_int_scalar} for the distance is equal to
\begin{equation}
  r\underset{\lambda\to 0}{=}2\sqrt{\frac{\lambda}{c}}\int\displaylimits_{0}^{1}dv\,\lsb
  \frac{v^2}{\sqrt{1-v^4}}+
  \frac{2}{3}\lambda\lb1+2\frac{U'(b,-1,0)}{U(b,-1,0)}\rb\frac{v^2(1-v^2)}{(1-v^4)^{3/2}}+
  O(\lambda^2)\rsb.
\end{equation}
For the energy integral~\eqref{en_int_scalar} we use the same form of the expansion,
\begin{equation}
  E\underset{\lambda\to0}{=}
  \frac{R^2}{\pi\alpha'}\sqrt{\frac{c}{\lambda}}\lsb\int\displaylimits_{0}^{1}
  \frac{dv}{v^2}\lb E_0+\lambda E_1+O(\lambda^2)\rb-U^{4/3}(b,-1,0)\rsb,
\end{equation}
with redefined expansion coefficients
\begin{equation}
  E_0\equiv U^{4/3}(b,-1,0)\lb\frac{1}{\sqrt{1-v^4}}-1\rb,
\end{equation}
\begin{equation}
  E_1\equiv \frac{2}{3}U^{4/3}(b,-1,0)\lb1+2\frac{U'(b,-1,0)}{U(b,-1,0)}\rb
  \frac{v^2(1-2v^4+v^2)}{(1-v^4)^{3/2}}.
\end{equation}
Then the remaining integrals can be computed just as they were for the vector case and this
results in~\eqref{en_small_lambda_scalar}.

For completeness, we present those properties of gamma-functions which were used
in the calculations above,
\begin{equation}
  \Gamma\lb\frac{1}{2}\rb=\sqrt{\pi},\quad
  \Gamma\lb\frac{3}{4}\rb\Gamma\lb\frac{1}{4}\rb=\pi\sqrt{2},\quad
  \Gamma(z+1)=z\Gamma(z),\quad
  \Gamma\lb-\frac{1}{2}\rb=-2\sqrt{\pi}.
\end{equation}

\bigskip

\section{Adjustments to earlier results}\label{adj_app}

In this Appendix, we review the adjustments to the results from~\cite{Andreev:2006ct} that
are required in order to make a meaningful comparison. The starting point is the
integral representation of the potential in terms of the parametric functions \(r(\lambda)\)
and \(E(\lambda)\) (counterparts to our~\eqref{r_int_def_2} and~\eqref{en_int_def_2}),
\begin{equation*}
  r=2\sqrt{\frac{\lambda}{c}}\int\displaylimits_{0}^{1}dv\,
  \frac{v^2e^{\lambda(1-v^2)/2}}{\sqrt{1-v^4e^{\lambda(1-v^2)}}},
\end{equation*}
\begin{equation*}
  E=\frac{R^2}{\pi\alpha'}\sqrt{\frac{c}{\lambda}}\lsb\int\displaylimits_{0}^{1}
  \frac{dv}{v^2}\lb\frac{e^{\lambda v^2/2}}{\sqrt{1-v^4e^{\lambda(1-v^2)}}}-1\rb-1\rsb.
\end{equation*}
The pertinent difference between~\cite{Andreev:2006ct} and the current work is the
difference in the definitions of the exponential background in the AdS$_5$ metric: \(e^{cz^2/2}\) vs. \(e^{2cz^2}\),
correspondingly. This implies that in order to use the above two formulae we have to
substitute \(c\to 4c\) and, consequently, \(\lambda\to 4\lambda\) due to its definition,
\(\lambda\equiv cz_0^2\). The integrals for the distance and energy change to the following ones,
\begin{equation}
  r=2\sqrt{\frac{\lambda}{c}}\int\displaylimits_{0}^{1}dv\,
  \frac{v^2e^{2\lambda(1-v^2)}}{\sqrt{1-v^4e^{4\lambda(1-v^2)}}},
\end{equation}
\begin{equation}
  E=\frac{R^2}{\pi\alpha'}\sqrt{\frac{c}{\lambda}}\lsb\int\displaylimits_{0}^{1}
  \frac{dv}{v^2}\lb\frac{e^{2\lambda v^2}}{\sqrt{1-v^4e^{4\lambda(1-v^2)}}}-1\rb-1\rsb.
\end{equation}
One can easily see that the last two formulae arise
from~\eqref{r_int_def_2} and~\eqref{en_int_def_2} in the limit of \(b\to 0\). Thus,
we can simply set \(b=0\) in~\eqref{pot_small_r} and obtain
\begin{equation}
  E\underset{r\to0}{=}\frac{R^2}{\alpha'}\lsb
  -\frac{1}{2\pi\rho^2}\frac{1}{r}+c\rho^2r\rsb.
\end{equation}
This result can be also verified by repeating the calculations of the
Section~\ref{small_r_sect} and Appendix~\ref{small_r_app}.

The situation with the large distance asymptotics is a little bit more complicated.
First, due to the substitution \(\lambda\to 4\lambda\), the limit of \(\lambda\to2\),
which is mapped to the limit of large distances in~\cite{Andreev:2006ct}, changes to
\(\lambda\to 1/2\). The second issue stems from the fact that the parameter \(\lambda\)
contributes to the overall coefficient of the potential. This is also the case in our
work: for example, the \(e^{2x}U^4(b,0,x)\) factor in~\eqref{pot_large_r} is due to
\(\lambda\)-containing coefficients. To clarify the second problem it makes sense to
reiterate the derivation procedure for the large distance potential once again
(this is also described in Ref.~\cite{Afonin:2021zdu}).

The main idea goes as follows: since the integrals for non-zero \(\lambda\) diverge
in the upper limit, the main contribution to their value comes from the integrands at
\(v=1\). Thus, we get
\begin{equation}\label{r_adj_AZ_interm}
  r= 2\sqrt{\frac{\lambda}{c}}\int\displaylimits_{0}^{1}
  \frac{dv}{\sqrt{(1-v)A(\lambda)+(v-1)^2B(\lambda)}},
\end{equation}
\begin{equation}\label{en_adj_AZ_interm}
  E=\frac{R^2}{\pi\alpha'}\sqrt{\frac{c}{\lambda}}e^{2\lambda}
  \int\displaylimits_{0}^{1}\frac{dv}{\sqrt{(1-v)A(\lambda)+(v-1)^2B(\lambda)}}.
\end{equation}
The new expression under the square root is the result of the series expansions and
since the original expressions under the square root were the same for \(r\) and \(E\),
the expansions above are also the same. Here \(A(\lambda)\) and \(B(\lambda)\) are some
functions, the precise form of which is not important for this discussion. Next, we
express the integral from~\eqref{r_adj_AZ_interm}, substitute it
into~\eqref{en_adj_AZ_interm} and arrive at
\begin{equation}
  E\underset{r\to\infty}{=}\frac{R^2}{2\pi\alpha'}\frac{c}{\lambda}e^{2\lambda}r.
\end{equation}
Finally, we substitute \(\lambda=1/2\) and obtain the large distance potential,
\begin{equation}
  E\underset{r\to\infty}{=}\frac{R^2}{\pi\alpha'}ecr.
\end{equation}
As an aside we would like to point out that in our reproduction of the results
of Ref.~\cite{Andreev:2006ct} we observed that the second function is actually equal to
\(B(\lambda)=9\lambda-2\lambda^2-6\) which differs from the corresponding expression in~\cite{Andreev:2006ct}. This
difference, however, does not affect the results because the relevant integral is simply canceled in the final formula.

\end{document}